\documentclass[iop]{emulateapj-rtx4}
\usepackage{natbib}
\usepackage{comment}
\usepackage{lscape}
\usepackage{amsmath}
\usepackage{hyperref,breakurl}
\usepackage{epsfig}

\shorttitle{RADIO AND MID-INFRARED PROPERTIES OF COMPACT STARBURSTS }
\shortauthors{MURPHY ET AL.}
\slugcomment{Submitted to ApJ October 30, 2012; Accepted February 20, 2013}

\begin{document}
\title{Radio and Mid-Infrared Properties of Compact Starbursts: Distancing Themselves from the Main Sequence}

\author{E.J.~Murphy\altaffilmark{1}, S.~Stierwalt\altaffilmark{2,3}, L.~Armus\altaffilmark{2}, J. J. Condon\altaffilmark{4}, and A.S.~Evans\altaffilmark{3,4}}
\altaffiltext{1}{Observatories of the Carnegie Institution for Science, 813 Santa Barbara Street, Pasadena, CA 91101, USA; emurphy@obs.carnegiescience.edu}
\altaffiltext{2}{{\it Spitzer Science Center,} California Institute of Technology, MC 314-6, Pasadena CA, 91125, USA}
\altaffiltext{3}{Department of Astronomy, University of Virginia, 530 McCormick Road, Charlottesville, VA 22904, USA}
\altaffiltext{4}{National Radio Astronomy Observatory, 520 Edgemont Road, Charlottesville, VA 22903, USA}

\begin{abstract}  
We investigate the relationship between 8.44\,GHz brightness temperatures and 1.4 to 8.44\,GHz radio spectral indices with 6.2\,$\mu$m polycyclic aromatic hydrocarbon (PAH) emission and 9.7\,$\mu$m silicate absorption features for a sample of 36 local luminous and ultra-luminous infrared galaxies.  
We find that galaxies having small 6.2\,$\mu$m PAH equivalent widths (EQWs), which signal the presence of weak PAH emission and/or an excess of very hot dust, also have flat spectral indices.  
The three active galactic nuclei (AGN) identified through their excessively large 8.44\,GHz brightness temperatures are also identified as AGN via their small 6.2\,$\mu$m PAH EQWs.  
We also find that the flattening of the radio spectrum increases with increasing silicate optical depth, 8.44\,GHz brightness temperature, and decreasing size of the radio source even after removing potential AGN, 
supporting the idea that compact starbursts show spectral flattening as the result of increased free-free absorption.  
These correlations additionally suggest that the dust obscuration in these galaxies must largely be coming from the vicinity of the compact starburst itself, and is not distributed throughout the (foreground) disk of the galaxy.
Finally, we investigate the location of these infrared-bright systems relative to the main sequence (star formation rate vs. stellar mass) of star-forming galaxies in the local universe.  
We find that the radio spectral indices of galaxies flattens with increasing distance above the main sequence, or in other words, with increasing specific star formation rate.  
This indicates that galaxies located above the main sequence, having high specific star formation rates, are typically compact starbursts hosting deeply embedded star formation that becomes more optically thick in the radio and infrared with increased distance above the main sequence.  

\end{abstract}

\section{Introduction}
Distinguishing the physical processes driving the energetics of infrared-bright galaxies is a non-trivial task.  
This is especially true for luminous infrared galaxies (LIRGs) whose infrared (IR; $8-1000\,\mu$m) luminosities exceed $L_{\rm IR} \ga 10^{11}\,L_{\sun}$.  
Immense column densities of dust towards the nuclei in these galaxies typically render traditional optical diagnostics (e.g. emission line flux ratios) difficult to interpret.
In the redshift range spanning $1 \la z \la 3$, such infrared-luminous galaxies appear to be much more common and dominate the star formation rate density, being an order of magnitude larger than today \citep[e.g.,][]{ce01,el05,kc07,bm11,ejm11a}.  

At these epochs, as well as today, there seems to be a distinction between the modes of star formation in infrared-bright sources that emerges by comparing their star formation rates and stellar masses.  
The majority of sources lie along a seemingly redshift dependent \citep{kgn07, de07, ed07a, mp09, gem10}
``main-sequence" where galaxies appear to be undergoing a mode of distributed star formation.  
A minority of sources, having compact star formation exhibited by elevated star formation rate surface densities, are instead found to lie systematically above the main sequence, having a high star formation rate per unit stellar mass \citep{de11}.   
The difference between these two populations is most likely related to the gas content and fueling of star formation in these systems, as illustrated by a similar separation of the same sources in the Schmidt law diagram  \citep{ms59,ks98}.  
Main sequence galaxies follow the classical Schmidt law while compact starbursts have star formation rates that are an order of magnitude larger for the same gas surface density \citep{ed10,rg10}.  
If the modes of star formation that are responsible for placing galaxies on and off the star-forming main sequence are the same at all redshifts, we should be able to use samples of local galaxies, for which we can obtain much more detailed information on small physical scales, to shed light on the detailed astrophysics of such distant systems.  

In the local universe, it is well-known that LIRGs and ultraluminous LIRGs (ULIRGs; $L_{\rm IR} \ga 10^{12}\,L_{\sun}$) appear to be undergoing an intense starburst phase.  
Within these systems are compact star-forming regions that have been been triggered predominantly through major mergers 
\citep[see e.g.,][]{lee87,lee88,lee89,lee90,dbs88a,dbs88b,twm96,sv95,sv97,sv02}.  
In some cases, the compact cores of ULIRGs have been resolved by high (i.e., sub-arcsecond) resolution, mid-infrared (i.e., $8-25\,\mu$m) imaging from the ground \citep[e.g.,][]{bts00}.  
However, given the large beams of typical space-based far-infrared (i.e., $\lambda \gtrsim 25\,\mu$m) telescopes, which are able to directly measure the bulk of the re-radiated energy from compact starbursts and active galactic nuclei (AGN) without suffering significantly from extinction, interferometric radio/mm/submm observations are currently the best means to {\it directly} resolve the sizes of these highly energetic, and often extremely compact, sources \citep[e.g.,][]{cjl06,ks09}.  

\citet{jc91} investigated the radio properties for a sample of 40 infrared-luminous galaxies included in the {\it IRAS} revised Bright Galaxies Sample \citep{bgs89,rbgs03}.  
Each source has a 60\,$\mu$m flux density larger than 5.24\,Jy and a dust temperature in the range of $60 \la T_{\rm d} \la 80$\,K, suggesting that their angular size at 60\,$\mu$m must be $\ga$0\farcs25.  
Thus, radio maps made at such a resolution should resolve the far-infrared source, leading \citet{jc91} to map the sample with 0\farcs25 resolution at 8.44\,GHz using the Very Large Array (VLA).  
In this paper we compare the results from \citet{jc91}, which measured 8.44\,GHz sizes and brightness temperatures of these compact, infrared-bright galaxies, with mid-infrared spectroscopic diagnostics taken as part of the Great Observatories All-Sky LIRG Survey \citep[GOALS;][]{lee09}.  
In doing this we attempt to see how the mid-infrared spectral properties compare with the physical conditions of the starbursts (e.g., sizes and brightness temperatures) as measured by the radio data.  

The paper is organized as follows:
In \S{2} we describe the existing radio and mid-infrared data used in the analysis.  
Then, in \S{3}, we compare various mid-infrared spectral and radio continuum properties of the sample, which are then discussed in \S{4}.  
Finally, we summarize our main conclusions in \S{5}.  

\section{Data and Analysis}
The galaxies in this sample are drawn from the {\it IRAS} revised Bright Galaxies Sample \citep{bgs89,rbgs03} having 60\,$\mu$m flux densities  larger than 5.24\,Jy and far-infrared (FIR; $42.5 - 122.5\,\mu$m) luminosities $\geq 10^{11.25}\,L_{\sun}$.  
The 40 systems that meet these criteria were imaged by \citet{jc91} at 8.44\,GHz with 0\farcs25 resolution, sometimes resolving multiple components.   
Of the resolved galaxies, radio spectral indices measured between 1.49 and 8.44\,GHz are available for 36 sources \citep{jc91}, which defines the sample used in the present analysis.  
We additionally make use of the derived 8.44\,GHz sizes and brightness temperatures available for 28 of the sources, which are all given in Table \ref{tbl-1}.  
As shown by \citet{jc91}, sources having 8.44\,GHz brightness temperatures $\ga10^{4.5}$\,K cannot be sustained by star formation alone, and must be powered by nuclear ``monsters" (AGN).  

The mid-infrared spectral properties were collected by the {\it Spitzer} Infrared Spectrograph \citep[IRS;][]{jh04} as part of GOALS \citep{lee09}, 
and are taken from \citet{ss13a, ss13b}.  
For 4 sources, IRS observations were either not taken (MCG\,-03-12-002) or the slit missed the nucleus of the galaxy (IRAS\,F01417+1651, IRAS\,F03359+1523, and IRAS\,F17132+5313).  
The total IR ($8-1000\,\mu$m) luminosities are taken from \citet{lee09}.  
However, for sources having multiple components as resolved by the IRS, the fractional IR luminosity from each component is determined by the fraction of resolved 24\,$\mu$m emission.  
We include 6.2$\mu$m PAH equivalent widths (EQWs), 9.7$\mu$m silicate optical depths, and silicate strengths \citep[see][]{ss13a, ss13b} in our analysis.  
The silicate strength at 9.7\,$\mu$m is defined as $s_{9.7\micron} =  \log(f_{9.7\micron}/C_{9.7\micron})$, where $f_{9.7\micron}$ is the measured flux density at the central wavelength of the absorption feature and $C_{9.7\micron}$ is the expected continuum level in the absence of the absorption feature.  
While the 0\farcs25 radio maps are at a resolution that is significantly higher than the mid-infrared IRS data (i.e., 3\farcs6), this should not affect our analysis or conclusions given that the central sources seem to dominate the luminosities of each system.  
Furthermore, the radio data themselves should be sensitive to extended emission on scales up to $\approx$5\arcsec, which is larger than the resolution of the IRS data.  

Stellar masses are taken from \citet{vu12}, who derived estimates by fitting optical through near-infrared spectral energy distributions, as well as assuming a fixed mass-to-light ratio and $H$-band flux densities, for both Salpeter \citep{salp55} and Chabrier \citep{chab03} IMFs.  
The masses adopted here are the mean of these estimates after first converting them to a Kroupa \citep{pk01} IMF using the following statistical relations as given by \citet{mbolz10}: \(\log{M_{*}({\rm Chab})} \approx \log{M_{*}({\rm Salp})} - 0.23\) and \(\log{M_{*}({\rm Krou})} \approx \log{M_{*}({\rm Chab})} + 0.04\).  
For sources having multiple components, we divided the total masses up by the proportions given by \citet{jh10}.  
These authors derived stellar masses using Two Micron All Sky Survey (2MASS) $K_{s}$-band (where possible) and IRAC 3.6\,$\mu$m photometry, along with the $z\sim0$ mass-to-light ratios presented by \citet{cgl08}.  
We prefer to use the mass estimates reported by \citet{vu12} as they have been more robustly determined.  
Each of these properties is given in Table \ref{tbl-1}.  
Infrared luminosities are converted into a star formation rate using the calibration of \citet{ejm11b}, which assumes a Kroupa IMF, such that,
\begin{equation}
\left(\frac{\rm SFR}{M_{\sun}\,{\rm yr^{-1}}}\right) = 1.48\times10^{-10}\left(\frac{L_{\rm IR}}{L_{\sun}}\right).
\end{equation}
We note that the coefficient here is 1.23 times larger, but within the scatter, of the empirically derived relation between IR and radio free-free star formation rates \citep{ejm12b}.

\begin{deluxetable*}{lccccccccc}
\tablecaption{Radio and Infrared Properties of the Sample Galaxies \label{tbl-1}}
\tabletypesize{\scriptsize}
\tablewidth{0pt}
\tablehead{
\colhead{Galaxy}  & \colhead{Dist.\tablenotemark{a}} & \colhead{$L_{\rm IR}/10^{11}$\tablenotemark{a}} & \colhead{$\alpha_{\rm 1.49\,GHz}^{\rm 8.44\,GHz}$\tablenotemark{b}} & \colhead{$\Omega/10^{-3}$\tablenotemark{b}} & \colhead{$T_{\rm b}/10^{4}$\tablenotemark{b}} & \colhead{$6.2\mu$m EQW\tablenotemark{c}} & \colhead{$\tau_{9.7\micron}$\tablenotemark{c}} & \colhead{$s_{9.7\micron}$\tablenotemark{c}} & \colhead{$M_{*}/10^{10}$\tablenotemark{d}}\\
\colhead{}  & \colhead{(Mpc)} & \colhead{($L_{\sun}$)} & \colhead{} & \colhead{(kpc$^{2}$)} & \colhead{(K)} & \colhead{($\mu$m)} & \colhead{} & \colhead{} & \colhead{($M_{\sun}$)}
}
\startdata
             NGC\,34&   84.1&   3.09&                      0.78&            33.96&             0.13&        0.45&             2.39&       -0.79&        2.08\\
       CGCG\,436-030&  134.0&   4.90&                      0.70&            89.38&             0.08&        0.35&             3.00&       -1.10&        1.20\\
   IRAS\,F01364-1042&  210.0&   7.08&                      0.42&            25.41&             0.63&        0.39&             5.14&       -1.27&        2.96\\
   IRAS\,F01417+1651&  119.0&   4.37&                      0.40&            15.64&             0.79&     \nodata&          \nodata&     \nodata&        1.17\\
            NGC\,695&  139.0&   4.79&    0.88 \tablenotemark{e}&          \nodata&          \nodata&        0.65&             0.57&        0.31&        9.77\\
           UGC\,2369&  136.0&   4.68&                      0.67&            12.76&             0.32&        0.57&             0.89&       -0.11&     \nodata\\
   IRAS\,F03359+1523&  152.0&   3.55&                      0.31&            10.48&             0.63&     \nodata&          \nodata&     \nodata&     \nodata\\
     MCG\,-03-12-002&  138.5&   3.24&                      0.65&             8.01&             0.63&     \nodata&          \nodata&     \nodata&     \nodata\\
           NGC\,1614&   67.8&   4.47&    1.05 \tablenotemark{e}&          \nodata&          \nodata&        0.61&             1.31&       -0.41&        2.47\\
   IRAS\,F05189-2524&  187.0&  14.45&                      0.52&            43.51&             0.40&        0.03&             1.39&       -0.29&        9.80\\
           NGC\,2623&   84.1&   4.37&                      0.58&            25.03&             0.40&        0.27&             3.15&       -1.12&        1.71\\
   IRAS\,F08572+3915&  264.0&  14.45&                      0.27&            15.95&             0.79&        0.03&             6.96&       -3.58&        1.59\\
       UGC\,04881\,E&  178.0&   3.63&                      0.69&            38.98&             0.25&        0.40&             2.00&       -0.81&        4.10\\
          UGC\,05101&  177.0&  10.23&                      0.59&            18.82&             3.16&        0.13&             3.65&       -0.78&        7.94\\
   IRAS\,F10173+0828&  224.0&   7.24&                      0.28&   $\leq$\,  7.63&   $\geq$\,  1.58&        0.35&             4.24&       -1.20&        2.55\\
   IRAS\,F10565+2448&  197.0&  12.02&                      0.68&          \nodata&          \nodata&        0.51&             2.13&       -0.75&        3.49\\
     MCG\,+07-23-019&  158.0&   4.17&                      0.55&            22.80&             0.40&        0.64&             2.31&       -0.55&     \nodata\\
   IRAS\,F11231+1456&  157.0&   4.37&                      0.82&          \nodata&          \nodata&        0.60&             1.16&       -0.22&        7.04\\
        NGC\,3690\,E&   50.8&   3.63&                      0.67&            11.55&             0.79&        0.38&             3.82&       -1.65&        0.98\\
   IRAS\,F12112+0305&  340.0&  22.91&                      0.47&            50.27&             0.79&        0.30&             3.42&       -1.24&        5.87\\
          UGC\,08058&  192.0&  37.15&                     -0.06&   $\leq$\,  5.51&   $\geq$\, 79.43&        0.01&     $\ga$\, 1.84&       -0.48&       30.63\\
         VV\,250a\,E&  142.0&   5.13&                      0.64&           125.76&             0.08&        0.63&             1.80&       -0.67&        0.90\\
          UGC\,08387&  110.0&   5.37&                      0.64&          \nodata&          \nodata&        0.62&             3.01&       -1.01&        2.30\\
        NGC\,5256\,S&  129.0&   3.09&    0.64 \tablenotemark{e}&          \nodata&          \nodata&        0.44&             1.50&       -0.47&        5.33\\
          UGC\,08696&  173.0&  16.22&                      0.63&            75.02&             0.63&        0.12&             4.24&       -1.37&        5.33\\
   IRAS\,F14348-1447&  387.0&  24.55&                      0.71&           149.04&             0.25&        0.25&             4.31&       -1.36&        6.79\\
         VV\,340a\,N&  157.0&   4.37&    0.60 \tablenotemark{e}&          \nodata&          \nodata&        0.58&             2.17&       -0.63&        4.61\\
   IRAS\,F15163+4255&  183.0&   8.32&                      0.78&            36.36&             0.25&        0.75&             1.06&        0.32&     \nodata\\
   IRAS\,F15250+3608&  254.0&  12.02&                      0.11&   $\leq$\,  7.51&   $\geq$\,  3.98&        0.03&             6.21&       -2.69&        2.16\\
          UGC\,09913&   87.9&  19.05&                      0.41&            10.09&             2.51&        0.17&             6.10&       -2.26&        4.48\\
        NGC\,6286\,S&   85.7&   1.62&    0.84 \tablenotemark{e}&          \nodata&          \nodata&        0.59&             2.28&       -0.40&        2.61\\
   IRAS\,F17132+5313&  232.0&   9.12&                      0.67&            46.64&             0.13&     \nodata&          \nodata&     \nodata&       11.14\\
   IRAS\,F22491-1808&  351.0&  15.85&                      0.41&   $\leq$\, 19.74&   $\geq$\,  0.79&        0.48&             3.23&       -1.04&        3.29\\
           NGC\,7469&   70.8&   4.47&    0.68 \tablenotemark{e}&             2.71&             1.26&        0.23&             0.70&        0.06&     \nodata\\
            IC\,5298&  119.0&   3.98&                      0.67&             5.78&             0.50&        0.12&             1.55&       -0.37&        4.15\\
            MRK\,331&   79.3&   3.16&                      0.66&             2.99&             0.79&        0.63&             1.59&       -0.35&        0.97  
\enddata
\tablenotetext{a}{Distances and total IR ($8-1000\,\mu$m) luminosities are taken from \citet{lee09}.}
\tablenotetext{b}{Radio data are taken from \citet{jc91}.}
\tablenotetext{c}{6.2\,$\mu$m PAH EQWs and silicate strengths ($s_{9.7\micron}$) are taken from \citep{ss13a} while silicate optical depths ($\tau_{9.7\micron}$) are take from \citet{ss13b}.}
\tablenotetext{d}{The mean mass estimate given in \citet{vu12} after converting to a Kroupa IMF.}
\tablenotetext{e}{Spectal indices are based on measurements at 1.49\,GHz and a frequency other than 8.44\,GHz \citep[see][]{jc91}.}
\end{deluxetable*}

\section{Results}
In this section we present the main findings by comparing the radio and mid-infrared properties of the sample.  

\subsection{Radio and Mid-Infrared Spectral Properties}

\begin{figure}
\epsscale{1.1}
\plotone{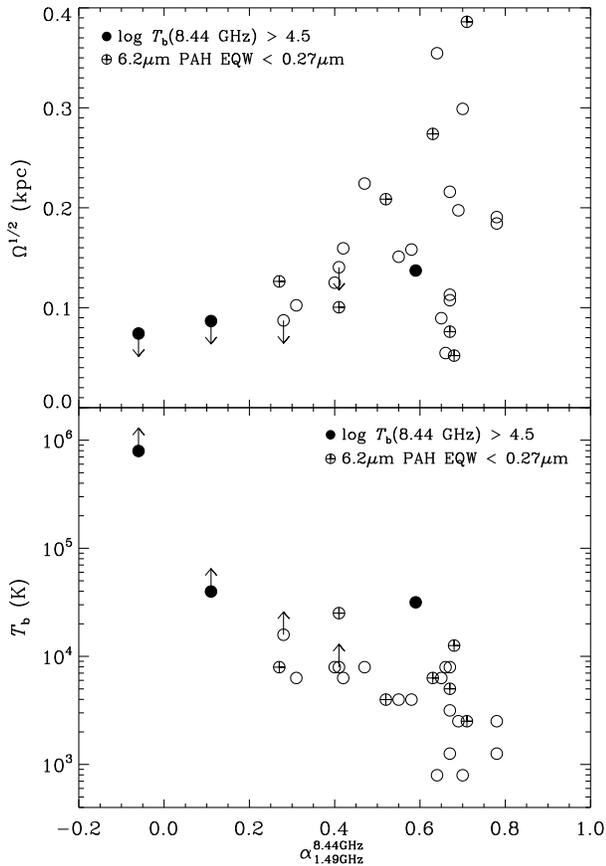}
\caption{Top: The linear 8.44\,GHz size of each source plotted against the radio spectral index (generally) measured between 1.49 and 8.44\,GHz (see Table \ref{tbl-1}).  
Bottom: The same as the top panel, except that the 8.44\,GHz brightness temperature is now plotted against the radio spectral spectral index.   
As already shown by \citet{jc91}, there is a general trend for sources having flat spectral indices to be small in size and have large 8.44\,GHz brightness temperatures.  
In both panels, sources with 8.44\,GHz brightness temperatures larger than $10^{4.5}\,$K, indicating the presence of an AGN, are identified by filled circles.  
We additionally identify those sources categorized as being AGN dominated by their low 6.2\,$\mu$m PAH EQW (circles with crosses); note, each of the sources with an 8.44\,GHz brightness temperature larger than $10^{4.5}\,$K also has a 6.2\,$\mu$m PAH EQW $<0.27\,\mu$m .  
}
\label{fig-spx-size}
\end{figure}

\begin{figure}
\epsscale{1.1}
\plotone{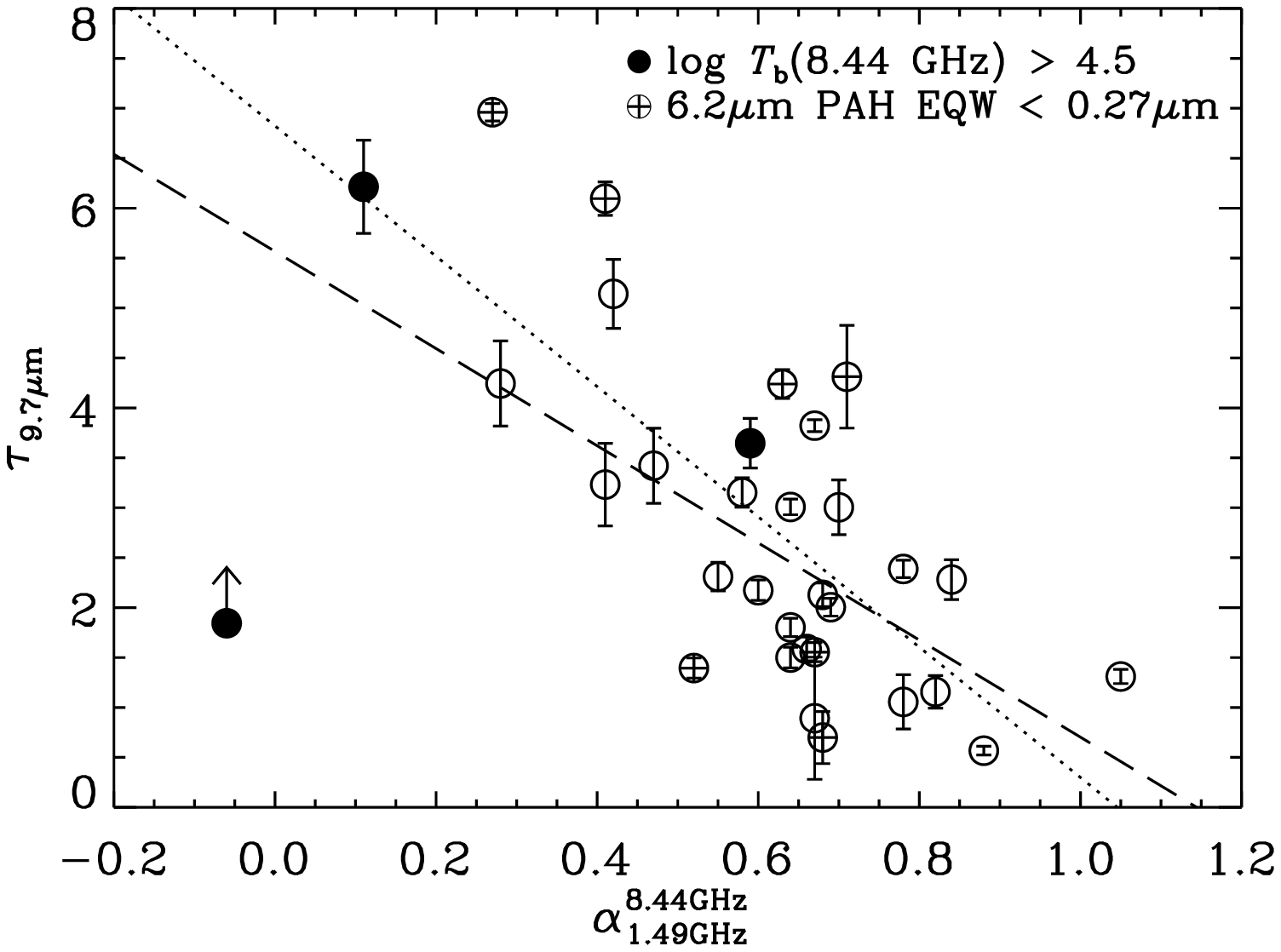}
\caption{The optical depth of the 9.7$\mu$m silicate feature plotted against the radio spectral index (generally) measured between 1.49 and 8.44\,GHz (see Table \ref{tbl-1}).  
Sources with 8.44\,GHz brightness temperatures larger than $10^{4.5}\,$K, indicating the presence of an AGN, are identified by filled circles.  
We additionally identify those sources categorized as being AGN dominated by their low 6.2\,$\mu$m PAH EQW (circles with crosses); note, each of the sources with an 8.44\,GHz brightness temperature larger than $10^{4.5}\,$K also has a 6.2\,$\mu$m PAH EQW $<0.27\,\mu$m.  
The dotted line is an ordinary least squares fit to all galaxies while the dashed line excludes sources identified as possible AGN.  
Both fits illustrate a general trend for sources to have decreasing silicate optical depths as their radio spectra steepen.  
The source to the lower left, having a negative radio spectral index, is UGC\,08058 (Mrk\,231) and is not included when fitting the data given that we only have a lower limit for its silicate optical depth.  
}
\label{fig-spx-tau}
\end{figure}

In the top panel of Figure \ref{fig-spx-size} we plot the 8.44\,GHz linear sizes of each source against their radio spectral index, typically measured between 1.49 and 8.44\,GHz (see Table \ref{tbl-1}).  
While there is a large range in size among galaxies having radio spectral indices steeper than $\alpha_{\rm 1.49GHz}^{\rm 8.44GHz} \gtrsim 0.5$, sources with $\alpha_{\rm 1.49GHz}^{\rm 8.44GHz} \lesssim 0.5$ all have small radio sizes.  
The mean and standard deviation in linear sizes among sources steeper and flatter than $\alpha_{\rm 1.49GHz}^{\rm 8.44GHz} = 0.5$ are $0.12\pm0.04$\,kpc and $0.18\pm0.10$\,kpc, respectively.  

By instead plotting the 8.44\,GHz brightness temperature against the radio spectral index (see bottom panel of Figure \ref{fig-spx-size}), 
there is a general trend as pointed out by \citet{jc91} in which more compact (i.e., higher surface brightness) sources tend to have flatter radio spectral indices, consistent with compact sources becoming optically thick at low (i.e., $\nu \la 5$\,GHz) radio frequencies.  
A similar trend is also seen when the radio spectral index is compared to the mid-infrared size of the starburst as measured from the {\it Spitzer} IRS data \citep{tds10}.  
Galaxies with flatter spectral indices tend to have smaller fractions of extended mid-infrared (13.2\,$\mu$m) continuum emission, again suggesting that the radio spectral index is related to the compactness of the starburst.  

In Figure \ref{fig-spx-tau} we plot the optical depth of the 9.7\,$\mu$m silicate feature against the radio spectral index.  
We find a correlation such that the silicate optical depth increases as the radio spectrum flattens. 
The only outlier in this trend is UGC\,08058 (Mrk\,231), which has a very small silicate optical depth given its extremely flat (inverted) radio spectrum.  
The shallow silicate optical depth in UGC\,08058 is most likely due to hot dust emission that fills in the absorption trough, thus the measured optical depth probably provides a lower limit.  
Additionally, its position in this diagram may also be complicated by the fact that the variable radio source in UGC\,08058 is small enough in size (i.e., $\la 1$\,pc) that its flat spectrum can be explained by synchrotron self-absorption \citep{jc91}.  
If we instead compare the radio spectral indices with the silicate strength, which does not rely on a model fit and measures the apparent silicate optical depth directly from the spectrum,  
we find a similar trend and UGC\,08058 remains an outlier.  

\begin{figure}
\epsscale{1.1}
\plotone{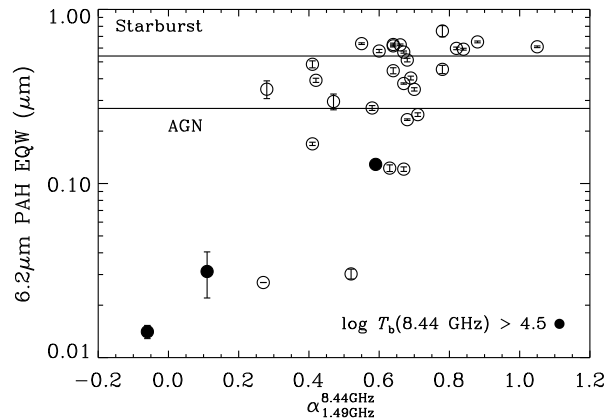}
\caption{The 6.2\,$\mu$m PAH EQW plotted against the radio spectral index (generally) measured between 1.49 and 8.44\,GHz (see Table \ref{tbl-1}).  
The horizontal lines indicate regions for which sources are thought to be AGN or star formation driven based on their measured 6.2\,$\mu$m PAH EQW.  
Sources having spectral indices steeper than $\alpha_{\rm 1.49GHz}^{\rm 8.44GHz} \gtrsim 0.6$ never appear to have extremely low 6.2$\mu$m PAH EQWs (i.e., $\la$0.1\,$\mu$m). 
Sources with 8.44\,GHz brightness temperatures larger than $10^{4.5}\,$K, indicating the presence of an AGN, are identified by filled circles.  
The source to the lower left, having a negative radio spectral index, is UGC\,08058 (Mrk\,231).  
}
\label{fig-spx-pah}
\end{figure}

\subsection{Identifying Potential AGN-Dominated Sources}
In Figure \ref{fig-spx-pah} we look at the 6.2\,$\mu$m PAH EQW, which measures the relative strength of the PAH flux with respect to the amount of hot dust emission at 6.2\,$\mu$m, versus radio spectral index.    
The 6.2\,$\mu$m PAH EQW has been used to indicate the presence of an AGN since they typically have very small PAH EQWs \citep[e.g.,][]{rg98,lee07}.
Specifically, starburst dominated systems appear to have 6.2\,$\mu$m PAH EQWs that are $\ga0.54\,\mu$m \citep{bb06}, while AGN have 6.2\,$\mu$m PAH EQWs are $\la 0.27\,\mu$m.  
We find that the galaxies having steep spectral indices (i.e., $\alpha_{\rm 1.49GHz}^{\rm 8.44GHz} \gtrsim 0.6$) never have extremely low 6.2$\mu$m PAH EQWs (i.e., $\la$0.1\,$\mu$m).  
Additionally, all 3 sources identified as AGN via their excessively large 8.44\,GHz brightness temperature sit in the PAH-defined AGN region.  

Using the 6.2\,$\mu$m PAH EQW to split the sources into starburst and AGN dominated systems, we find that the mean radio spectral index for starburst dominated systems is 0.74, with a standard deviation of 0.15.    
For the AGN dominated systems, we find a significantly flatter mean spectral index of 0.45, albeit with a much larger scatter of 0.27.  

\section{Discussion}
Taking archival high resolution 8.44\,GHz data from the literature, we have compared the radio and mid-infrared spectral properties for a sample of local (U)LIRGs.  
We now use these results to investigate how the dust is distributed in these powerful, infrared-bright galaxies.    
We additionally look at how the radio properties of these sources relate to their location in the star formation rate -- stellar mass plane of star-forming galaxies.  

\subsection{The Dust Distribution in Compact Starbursts}  
In Figure \ref{fig-spx-size} we have shown that the 8.44\,GHz radio sizes (top panel) and brightness temperatures (bottom panel) of sources decrease and increase, respectively, as a galaxy's radio spectrum flattens. 
While it may not be too surprising to find compact, flat-spectrum AGN, which indeed make up the two sources having the flattest radio spectra, we find that these trends persist even after identifying potential AGN-dominated sources, as indicated by  low 6.2\,$\mu$m PAH EQW (i.e., $< 0.27\,\mu$m) or high 8.44\,GHz brightness temperature [i.e., $T_{\rm b}(8.44\,{\rm GHz}) > 10^{4.5}$\,K].  

Additionally, in Figure \ref{fig-spx-tau}, we find a  trend such that the silicate optical depth increases with the flattening of the radio spectral index.  
Sources potentially dominated by AGN, as indicated by their low 6.2\,$\mu$m PAH EQW (i.e., $< 0.27\,\mu$m) or high 8.44\,GHz brightness temperature [i.e., $T_{\rm b}(8.44\,{\rm GHz}) > 10^{4.5}$\,K] are identified.  
An ordinary least squares fit to the data, excluding UGC\,08058 for which we only have a lower limit on its silicate optical depth,  
is shown as a dotted line and yields the following relation:
\begin{equation}
\label{eq-d_ms-spx}
\tau_{9.7\micron} = (-6.5\pm 1.0) \alpha + (6.8\pm 0.7).
\end{equation}
The fact that we see a trend with flat spectrum sources having larger silicate optical depths lends support to the idea that starbursts show spectral flattening as the result of increased free-free absorption arising from more deeply embedded star formation.  
In fact, if we instead limit the fit to non-AGN sources, as indicated by either a low  6.2\,$\mu$m PAH EQW or a high 8.44\,GHz brightness temperature, the trend actually tightens; 
the residual dispersion about the fit (see the dashed line in Figure \ref{fig-spx-tau}), whose coefficients are consistent within errors [\(\tau_{9.7\micron} = (-4.9\pm 1.0) \alpha + (5.6\pm 0.6)\)], is 30\% smaller than when fitting the entire sample.  
The increased dispersion by including AGN may not be surprising given that, unlike optically-thick starbursts, flat-spectrum AGN need not be heavily obscured. 

Since the radio observations are not significantly affected by dust extinction, we can assume that the radio data are able to penetrate the starburst core itself.  
\citet{jc91} showed that the spectral flattening observed in the radio data can be explained by the starburst becoming increasingly compact and energetic.  
Thus, the correlation between the spectral flattening and increased silicate optical depth (Figure \ref{fig-spx-tau}) and the correlations between the spectral flattening and the radio extent and brightness temperature (Figure \ref{fig-spx-size}) most likely indicates that the dust obscuration 
in these sources occurs in the vicinity of the compact starburst itself, and not by extended dust located in the foreground galaxy disks.

\begin{figure}
\epsscale{1.1}
\plotone{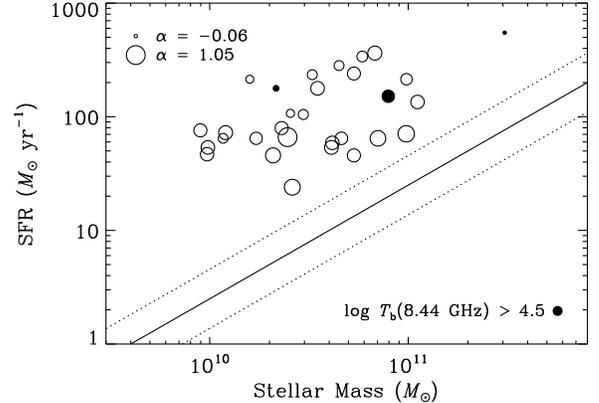}
\caption{The infrared-derived star formation rate of each galaxy plotted against its stellar mass.  
The size of each plotting symbol is a function of the galaxy's radio spectral index, with smaller sizes indicating flatter spectral indices.  
The solid and dashed lines indicate the main-sequence and 1$\sigma$ (0.26\,dex) dispersion for 9\,$\mu$m detected AKARI galaxies at $z\sim0$ as measured by \citet[][${\rm SFR}/M_{*} \sim 0.25\,{\rm Gyr}^{-1}$]{de11}.  
Each of the infrared-luminous galaxies in this sample is located well above the main-sequence relation.  
Sources with 8.44\,GHz brightness temperatures larger than $10^{4.5}\,$K, indicating the presence of an AGN, are identified by filled circles.  
}
\label{fig-sfr-mass}
\end{figure}

\subsection{Radio Spectral Indices and the Main Sequence of Star-Forming Galaxies}
\label{sec-spx-ssfr}
In Figure \ref{fig-sfr-mass} we plot the $z\sim0$ main sequence relation as given by \citet{de11}, which is established using a local sample of galaxies detected at 9\,$\mu$m with the AKARI/Infrared Camera \citep[IRC;][]{to07}.  
Most galaxies lie along a linear relation between star formation rate and stellar mass, such that the specific star formation rate (${\rm sSFR = SFR}/M_{*}$) is constant, being ${\rm sSFR} \sim 0.25\,{\rm Gyr}^{-1}$ with a dispersion (standard deviation) that is a factor of 1.82.  
We note that the $z\sim0$ main-sequence relation reported by \citet{de07} has a slope that is slightly flatter than unity, being 0.77.  
If we instead adopt this relation for the $z\sim0$ main sequence, our conclusions are not strongly affected. 
Like \citet{de11}, we decide to chose the main-sequence relation having a slope of unity as this is consistent with findings at $z\sim1$ \citep{de07}, $z\sim2$ \citep{mp09}, $z\sim3$ \citep{gem10}, and $z\sim4$ \citep{ed09}.  

We overplot our sample galaxies and find that each source lies more than 1$\sigma$ away from the main sequence, as expected for starbursting galaxies.  
Sources having the flattest spectral indices tend to be furthest from the main sequence.  

\begin{figure}[ht!]
\epsscale{1.1}
\plotone{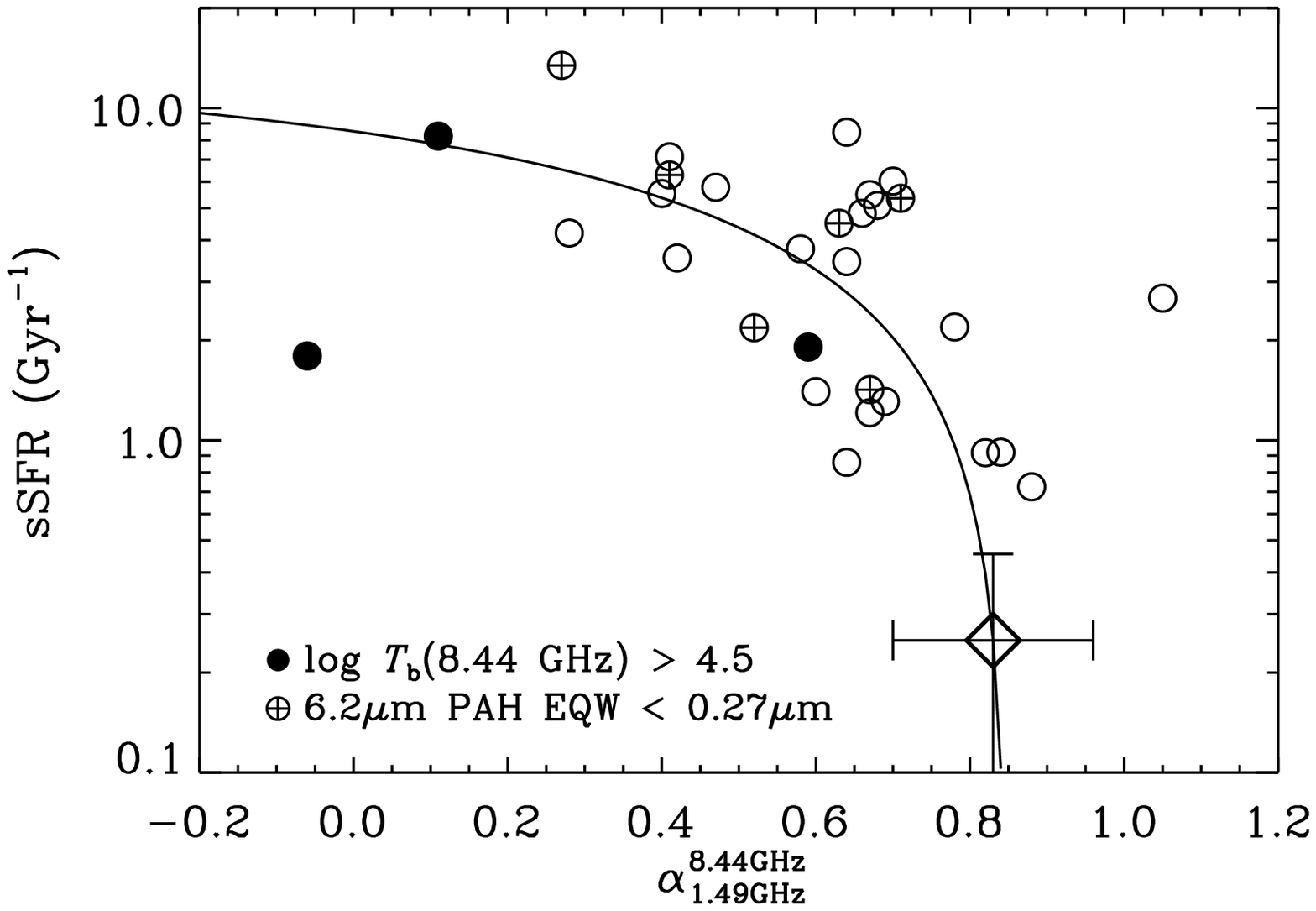}
\caption{The specific star formation rate of each galaxy plotted as a function of radio spectral index (generally) measured between 1.49 and 8.44\,GHz (see Table \ref{tbl-1}).  
Sources with 8.44\,GHz brightness temperatures larger than $10^{4.5}\,$K, indicating the presence of an AGN, are identified by filled circles.  
We additionally identify those sources categorized as being AGN dominated by their low 6.2\,$\mu$m PAH EQW (circles with crosses); note, each of the sources with an 8.44\,GHz brightness temperature larger than $10^{4.5}\,$K also has a 6.2\,$\mu$m PAH EQW $<0.27\,\mu$m .  
The solid line is a fit to the non-AGN sources, as identified by either their low 6.2\,$\mu$m PAH EQW or high 8.44\,GHz brightness temperature, of the form 
\({\rm sSFR} = \eta(1-e^{\alpha - \alpha_{\rm MS}}) + {\rm sSFR_{MS}}\,e^{\alpha - \alpha_{\rm MS}}\)
(see \S\ref{sec-spx-ssfr}), which is forced to fit through the typical radio spectral index ($\alpha_{\rm MS} \sim 0.8$) and specific star formation rate (${\rm sSFR_{MS}} \sim 0.25\,{\rm Gyr}^{-1}$) of $z\sim0$ star-forming galaxies (diamond).  
The corresponding error bars indicate the 1$\sigma$ dispersions of these parameters (i.e., $\sigma_{\alpha_{\rm MS}} = 0.13$ and $\sigma_{\rm sSFR_{\rm MS}} = 0.26$\,dex) 
}
\label{fig-ssfr-spx}
\end{figure}

In Figure \ref{fig-ssfr-spx} we investigate the relation between the distance of these galaxies from the main sequence versus the compactness of the starburst as measured by their radio spectral index.  
Since we have adopted a constant specific star formation rate to define the $z\sim0$ main sequence, the distance from the main sequence is given by the specific star formation rate itself.  
We find a general trend where the specific star formation rate among the sample increases as the radio spectral index flattens.  
While a general trend does exist, 
the entire sample lies well above the $z\sim0$ main sequence.  
The minimum and mean specific star formation rates among the sample are a factor of $\sim$2.5 and 16 times larger than the $z\sim0$ main sequence value, respectively.    
However, many sources have radio spectral indices consistent with normal star-forming galaxies \citep[i.e., $\alpha \sim 0.8$;][]{jc92}, which comfortably lie on the main sequence.     
Thus, it seems that a flat radio spectral index does indicate that a source is above the main sequence, but galaxies having a ``normal" star-forming radio spectra (i.e., $\alpha \sim 0.8$) are not necessarily on the main sequence.  


Since we know that typical $z \sim 0$ star-forming galaxies lying on the main sequence have radio spectral indices near $\alpha \sim 0.8$, it may be more appropriate to fit the data with something other than an ordinary least squares fit that passes through this normalization point.  
The radio brightness temperature, which is a measure of the surface brightness or ``compactness" of a source, is given by the expression 
\(T_{\rm b} = T_{\rm e}(1-e^{-\tau_{\rm ff}}) + T_{\rm b}(0)e^{-\tau_{\rm ff}}\), 
where $T_{\rm e}$ is the thermal electron temperature, $\tau_{\rm ff}$ is the free-free optical depth, and $T_{\rm b}(0)$ is the initial brightness temperature.  
The specific star formation rate has been shown to be related to the compactness (brightness temperature) of a source \citep[e.g.,][]{de11}, and in Figure \ref{fig-spx-tau} we have found a relation between the radio spectral index and the optical depth of the 9.7\,$\mu$m silicate feature.  
Thus, let us assume that the specific star formation rate is related to the radio spectral index with a function having the form of 
\({\rm sSFR} = \eta(1-e^{\alpha - \alpha_{\rm MS}}) + {\rm sSFR_{MS}}\,e^{\alpha - \alpha_{\rm MS}}\), 
where $\alpha_{\rm MS}$ and ${\rm sSFR_{MS}}$ are the typical radio spectral indices and specific star formation rates for galaxies on the $z \sim 0$ main sequence, respectively, and $\eta$ is a normalization constant related to the maximum specific star formation rate, which is determined by the fit to the data.  

Taking $\alpha_{\rm MS} = 0.83$ with a standard deviation of $\sigma_{\alpha_{\rm MS}} = 0.13$ \citep{nkw97} and ${\rm sSFR} \sim 0.25\,{\rm Gyr}^{-1}$ with a standard deviation that is a factor of 1.82 \citep{de11}, we find that sources not identified as harboring AGN by either a low 6.2\,$\mu$m PAH EQW (i.e., $< 0.27\,\mu$m) or high 8.44\,GHz brightness temperature [i.e., $T_{\rm b}(8.44\,{\rm GHz}) > 10^{4.5}$\,K] are best fit by 
\begin{equation}
\label{eq-ssfr-spx-fix}
\left(\frac{\rm sSFR}{\rm Gyr^{-1}} \right) = (14.9\pm2.9) (1-e^{\alpha - 0.83})+ \left( \frac{0.25}{\rm Gyr^{-1}}\right)e^{\alpha - 0.83},   
\end{equation}
where the error on $\eta$ is estimated by the standard deviation of the $\chi^{2}$ distribution from the fit.
We show this fit to the data in Figure \ref{fig-ssfr-spx}, along with the $z\sim0$ normalization point (diamond) and 1$\sigma$ dispersions.  
If we instead only exclude those sources identified as AGN by their excessively large 8.44\,GHz brightness temperatures, given that it is unclear whether the small 6.2\,$\mu$m PAH EQW is completely the result of an embedded AGN, we obtain a slightly larger value for $\eta$, being $\approx17.1\pm4.3$.  

\section{Conclusions}
We have compared the mid-infrared spectral properties of local infrared-luminous systems with high resolution radio measurements in the literature, specifically comparing 8.44\,GHz brightness temperatures and 1.4 to 8.44\,GHz radio spectral indices with 6.2\,$\mu$m polycyclic aromatic hydrocarbon (PAH) and 9.7\,$\mu$m silicate absorption features.
In doing this, we have also investigated the relations between the main-sequence of star-forming galaxies in the local universe, as defined in the star formation rate vs. stellar mass plane, and their compactness as indicated by their radio spectral indices.  
Our two main conclusions can be summarized as follows:

\begin{itemize}

\item{
We find that the flattening of the radio spectrum increases with increasing silicate optical depth, 8.44\,GHz brightness temperature, and decreasing size of the radio source even after removing potential AGN, 
supporting the idea that compact starbursts show spectral flattening as the result of increased free-free absorption.  
This trend additionally suggests that dust obscuration must largely be coming from the vicinity of the compact starburst itself and not the foreground galaxy disk.  }

\item{
We observe a trend such that the radio spectral index of galaxies flattens with increasing distance above the main sequence (i.e. with increasing specific star formation rate).  
This suggests that starbursts become more compact and deeply embedded with increasing specific star formation rate, moving them further away from the main sequence of star-forming galaxies.}
\end{itemize}

With the Karl G. Jansky VLA now online, we have reached a point where it is feasible to significantly improve upon the type of analysis presented in this paper by conducting higher frequency radio continuum surveys (e.g., $\ga$10\,GHz) of local starbursts, probing $\la$100\,pc scales \citep[e.g.,][]{akl11b}.  
For instance, by having resolved sizes along with better spectral coverage, to identify the cutoff frequency at which the free-free optical depth equals unity, one can begin to characterize additional physical characteristics of the starbursts such as their emission measures and electron densities.  
Additionally, we are likely to see a new generation of high-frequency, deep-field radio surveys that compliment existing 1.4\,GHz data, while providing sub-arcsecond resolution of distant dusty starbursts.  
Such surveys will need to rely on investigations such as these to interpret their observations.  

\acknowledgements
We thank the anonymous referee for useful comments that helped to significantly improve the content and presentation of this paper.
We thank J.H.~Howell, T.~D\'{i}az-Santos, and V. Charmandaris for useful discussions.  
The National Radio Astronomy Observatory is a facility of the National Science Foundation operated under cooperative agreement by Associated Universities, Inc.
This work is based in part on observations made with the {\it Spitzer} Space Telescope, which is operated by the Jet Propulsion Laboratory, California Institute of Technology under a contract with NASA.  


\bibliography{/Users/emurphy/libs/bibtexref/master_ref}

\begin{thebibliography}{53}
\expandafter\ifx\csname natexlab\endcsname\relax\def\natexlab#1{#1}\fi

\bibitem[{{Armus} {et~al.}(1987){Armus}, {Heckman}, \& {Miley}}]{lee87}
{Armus}, L., {Heckman}, T., \& {Miley}, G. 1987, \aj, 94, 831

\bibitem[{{Armus} {et~al.}(1988){Armus}, {Heckman}, \& {Miley}}]{lee88}
{Armus}, L., {Heckman}, T.~M., \& {Miley}, G.~K. 1988, \apjl, 326, L45

\bibitem[{{Armus} {et~al.}(1989){Armus}, {Heckman}, \& {Miley}}]{lee89}
---. 1989, \apj, 347, 727

\bibitem[{{Armus} {et~al.}(1990){Armus}, {Heckman}, \& {Miley}}]{lee90}
---. 1990, \apj, 364, 471

\bibitem[{{Armus} {et~al.}(2007){Armus}, {Charmandaris}, {Bernard-Salas},
  {Spoon}, {Marshall}, {Higdon}, {Desai}, {Teplitz}, {Hao}, {Devost}, {Brandl},
  {Wu}, {Sloan}, {Soifer}, {Houck}, \& {Herter}}]{lee07}
{Armus}, L., {Charmandaris}, V., {Bernard-Salas}, J., {et~al.} 2007, \apj, 656,
  148

\bibitem[{{Armus} {et~al.}(2009){Armus}, {Mazzarella}, {Evans}, {Surace},
  {Sanders}, {Iwasawa}, {Frayer}, {Howell}, {Chan}, {Petric}, {Vavilkin},
  {Kim}, {Haan}, {Inami}, {Murphy}, {Appleton}, {Barnes}, {Bothun}, {Bridge},
  {Charmandaris}, {Jensen}, {Kewley}, {Lord}, {Madore}, {Marshall},
  {Melbourne}, {Rich}, {Satyapal}, {Schulz}, {Spoon}, {Sturm}, {U}, {Veilleux},
  \& {Xu}}]{lee09}
{Armus}, L., {Mazzarella}, J.~M., {Evans}, A.~S., {et~al.} 2009, \pasp, 121,
  559

\bibitem[{{Bolzonella} {et~al.}(2010){Bolzonella}, {Kova{\v c}}, {Pozzetti},
  {Zucca}, {Cucciati}, {Lilly}, {Peng}, {Iovino}, {Zamorani}, {Vergani},
  {Tasca}, {Lamareille}, {Oesch}, {Caputi}, {Kampczyk}, {Bardelli}, {Maier},
  {Abbas}, {Knobel}, {Scodeggio}, {Carollo}, {Contini}, {Kneib}, {Le
  F{\`e}vre}, {Mainieri}, {Renzini}, {Bongiorno}, {Coppa}, {de la Torre}, {de
  Ravel}, {Franzetti}, {Garilli}, {Le Borgne}, {Le Brun}, {Mignoli},
  {Pell{\'o}}, {Perez-Montero}, {Ricciardelli}, {Silverman}, {Tanaka},
  {Tresse}, {Bottini}, {Cappi}, {Cassata}, {Cimatti}, {Guzzo}, {Koekemoer},
  {Leauthaud}, {Maccagni}, {Marinoni}, {McCracken}, {Memeo}, {Meneux},
  {Porciani}, {Scaramella}, {Aussel}, {Capak}, {Halliday}, {Ilbert},
  {Kartaltepe}, {Salvato}, {Sanders}, {Scarlata}, {Scoville}, {Taniguchi}, \&
  {Thompson}}]{mbolz10}
{Bolzonella}, M., {Kova{\v c}}, K., {Pozzetti}, L., {et~al.} 2010, \aap, 524,
  A76

\bibitem[{{Brandl} {et~al.}(2006){Brandl}, {Bernard-Salas}, {Spoon}, {Devost},
  {Sloan}, {Guilles}, {Wu}, {Houck}, {Weedman}, {Armus}, {Appleton}, {Soifer},
  {Charmandaris}, {Hao}, {Higdon}, {Marshall}, \& {Herter}}]{bb06}
{Brandl}, B.~R., {Bernard-Salas}, J., {Spoon}, H.~W.~W., {et~al.} 2006, \apj,
  653, 1129

\bibitem[{{Caputi} {et~al.}(2007){Caputi}, {Lagache}, {Yan}, {Dole},
  {Bavouzet}, {Le Floc'h}, {Choi}, {Helou}, \& {Reddy}}]{kc07}
{Caputi}, K.~I., {Lagache}, G., {Yan}, L., {et~al.} 2007, \apj, 660, 97

\bibitem[{{Chabrier}(2003)}]{chab03}
{Chabrier}, G. 2003, \pasp, 115, 763

\bibitem[{{Chary} \& {Elbaz}(2001)}]{ce01}
{Chary}, R., \& {Elbaz}, D. 2001, \apj, 556, 562

\bibitem[{{Condon}(1992)}]{jc92}
{Condon}, J.~J. 1992, \araa, 30, 575

\bibitem[{{Condon} {et~al.}(1991){Condon}, {Huang}, {Yin}, \& {Thuan}}]{jc91}
{Condon}, J.~J., {Huang}, Z.-P., {Yin}, Q.~F., \& {Thuan}, T.~X. 1991, \apj,
  378, 65

\bibitem[{{Daddi} {et~al.}(2007){Daddi}, {Dickinson}, {Morrison}, {Chary},
  {Cimatti}, {Elbaz}, {Frayer}, {Renzini}, {Pope}, {Alexander}, {Bauer},
  {Giavalisco}, {Huynh}, {Kurk}, \& {Mignoli}}]{ed07a}
{Daddi}, E., {Dickinson}, M., {Morrison}, G., {et~al.} 2007, \apj, 670, 156

\bibitem[{{Daddi} {et~al.}(2009){Daddi}, {Dannerbauer}, {Stern}, {Dickinson},
  {Morrison}, {Elbaz}, {Giavalisco}, {Mancini}, {Pope}, \& {Spinrad}}]{ed09}
{Daddi}, E., {Dannerbauer}, H., {Stern}, D., {et~al.} 2009, \apj, 694, 1517

\bibitem[{{Daddi} {et~al.}(2010){Daddi}, {Elbaz}, {Walter}, {Bournaud},
  {Salmi}, {Carilli}, {Dannerbauer}, {Dickinson}, {Monaco}, \&
  {Riechers}}]{ed10}
{Daddi}, E., {Elbaz}, D., {Walter}, F., {et~al.} 2010, \apjl, 714, L118

\bibitem[{{D{\'{\i}}az-Santos} {et~al.}(2010){D{\'{\i}}az-Santos},
  {Charmandaris}, {Armus}, {Petric}, {Howell}, {Murphy}, {Mazzarella},
  {Veilleux}, {Bothun}, {Inami}, {Appleton}, {Evans}, {Haan}, {Marshall},
  {Sanders}, {Stierwalt}, \& {Surace}}]{tds10}
{D{\'{\i}}az-Santos}, T., {Charmandaris}, V., {Armus}, L., {et~al.} 2010, \apj,
  723, 993

\bibitem[{{Elbaz} {et~al.}(2007){Elbaz}, {Daddi}, {Le Borgne}, {Dickinson},
  {Alexander}, {Chary}, {Starck}, {Brandt}, {Kitzbichler}, {MacDonald},
  {Nonino}, {Popesso}, {Stern}, \& {Vanzella}}]{de07}
{Elbaz}, D., {Daddi}, E., {Le Borgne}, D., {et~al.} 2007, \aap, 468, 33

\bibitem[{{Elbaz} {et~al.}(2011){Elbaz}, {Dickinson}, {Hwang},
  {D{\'{\i}}az-Santos}, {Magdis}, {Magnelli}, {Le Borgne}, {Galliano},
  {Pannella}, {Chanial}, {Armus}, {Charmandaris}, {Daddi}, {Aussel}, {Popesso},
  {Kartaltepe}, {Altieri}, {Valtchanov}, {Coia}, {Dannerbauer}, {Dasyra},
  {Leiton}, {Mazzarella}, {Alexander}, {Buat}, {Burgarella}, {Chary}, {Gilli},
  {Ivison}, {Juneau}, {Le Floc'h}, {Lutz}, {Morrison}, {Mullaney}, {Murphy},
  {Pope}, {Scott}, {Brodwin}, {Calzetti}, {Cesarsky}, {Charlot}, {Dole},
  {Eisenhardt}, {Ferguson}, {F{\"o}rster Schreiber}, {Frayer}, {Giavalisco},
  {Huynh}, {Koekemoer}, {Papovich}, {Reddy}, {Surace}, {Teplitz}, {Yun}, \&
  {Wilson}}]{de11}
{Elbaz}, D., {Dickinson}, M., {Hwang}, H.~S., {et~al.} 2011, \aap, 533, A119

\bibitem[{{Genzel} {et~al.}(1998){Genzel}, {Lutz}, {Sturm}, {Egami}, {Kunze},
  {Moorwood}, {Rigopoulou}, {Spoon}, {Sternberg}, {Tacconi-Garman}, {Tacconi},
  \& {Thatte}}]{rg98}
{Genzel}, R., {Lutz}, D., {Sturm}, E., {et~al.} 1998, \apj, 498, 579

\bibitem[{{Genzel} {et~al.}(2010){Genzel}, {Tacconi}, {Gracia-Carpio},
  {Sternberg}, {Cooper}, {Shapiro}, {Bolatto}, {Bouch{\'e}}, {Bournaud},
  {Burkert}, {Combes}, {Comerford}, {Cox}, {Davis}, {Schreiber},
  {Garcia-Burillo}, {Lutz}, {Naab}, {Neri}, {Omont}, {Shapley}, \&
  {Weiner}}]{rg10}
{Genzel}, R., {Tacconi}, L.~J., {Gracia-Carpio}, J., {et~al.} 2010, \mnras,
  407, 2091

\bibitem[{{Houck} {et~al.}(2004){Houck}, {Roellig}, {van Cleve}, {Forrest},
  {Herter}, {Lawrence}, {Matthews}, {Reitsema}, {Soifer}, {Watson}, {Weedman},
  {Huisjen}, {Troeltzsch}, {Barry}, {Bernard-Salas}, {Blacken}, {Brandl},
  {Charmandaris}, {Devost}, {Gull}, {Hall}, {Henderson}, {Higdon}, {Pirger},
  {Schoenwald}, {Sloan}, {Uchida}, {Appleton}, {Armus}, {Burgdorf},
  {Fajardo-Acosta}, {Grillmair}, {Ingalls}, {Morris}, \& {Teplitz}}]{jh04}
{Houck}, J.~R., {Roellig}, T.~L., {van Cleve}, J., {et~al.} 2004, \apjs, 154,
  18

\bibitem[{{Howell} {et~al.}(2010){Howell}, {Armus}, {Mazzarella}, {Evans},
  {Surace}, {Sanders}, {Petric}, {Appleton}, {Bothun}, {Bridge}, {Chan},
  {Charmandaris}, {Frayer}, {Haan}, {Inami}, {Kim}, {Lord}, {Madore},
  {Melbourne}, {Schulz}, {U}, {Vavilkin}, {Veilleux}, \& {Xu}}]{jh10}
{Howell}, J.~H., {Armus}, L., {Mazzarella}, J.~M., {et~al.} 2010, \apj, 715,
  572

\bibitem[{{Kennicutt}(1998)}]{ks98}
{Kennicutt}, Jr., R.~C. 1998, \apj, 498, 541

\bibitem[{{Kroupa}(2001)}]{pk01}
{Kroupa}, P. 2001, \mnras, 322, 231

\bibitem[{{Lacey} {et~al.}(2008){Lacey}, {Baugh}, {Frenk}, {Silva}, {Granato},
  \& {Bressan}}]{cgl08}
{Lacey}, C.~G., {Baugh}, C.~M., {Frenk}, C.~S., {et~al.} 2008, \mnras, 385,
  1155

\bibitem[{{Le Floc'h} {et~al.}(2005){Le Floc'h}, {Papovich}, {Dole}, {Bell},
  {Lagache}, {Rieke}, {Egami}, {P{\'e}rez-Gonz{\'a}lez}, {Alonso-Herrero},
  {Rieke}, {Blaylock}, {Engelbracht}, {Gordon}, {Hines}, {Misselt}, {Morrison},
  \& {Mould}}]{el05}
{Le Floc'h}, E., {Papovich}, C., {Dole}, H., {et~al.} 2005, \apj, 632, 169

\bibitem[{{Leroy} {et~al.}(2011){Leroy}, {Evans}, {Momjian}, {Murphy}, {Ott},
  {Armus}, {Condon}, {Haan}, {Mazzarella}, {Meier}, {Privon}, {Schinnerer},
  {Surace}, \& {Walter}}]{akl11b}
{Leroy}, A.~K., {Evans}, A.~S., {Momjian}, E., {et~al.} 2011, \apjl, 739, L25

\bibitem[{{Lonsdale} {et~al.}(2006){Lonsdale}, {Diamond}, {Thrall}, {Smith}, \&
  {Lonsdale}}]{cjl06}
{Lonsdale}, C.~J., {Diamond}, P.~J., {Thrall}, H., {Smith}, H.~E., \&
  {Lonsdale}, C.~J. 2006, \apj, 647, 185

\bibitem[{{Magdis} {et~al.}(2010){Magdis}, {Rigopoulou}, {Huang}, \&
  {Fazio}}]{gem10}
{Magdis}, G.~E., {Rigopoulou}, D., {Huang}, J.-S., \& {Fazio}, G.~G. 2010,
  \mnras, 401, 1521

\bibitem[{{Magnelli} {et~al.}(2011){Magnelli}, {Elbaz}, {Chary}, {Dickinson},
  {Le Borgne}, {Frayer}, \& {Willmer}}]{bm11}
{Magnelli}, B., {Elbaz}, D., {Chary}, R.~R., {et~al.} 2011, \aap, 528, A35+

\bibitem[{{Murphy} {et~al.}(2011{\natexlab{a}}){Murphy}, {Chary}, {Dickinson},
  {Pope}, {Frayer}, \& {Lin}}]{ejm11a}
{Murphy}, E.~J., {Chary}, R.-R., {Dickinson}, M., {et~al.} 2011{\natexlab{a}},
  \apj, 732, 126

\bibitem[{{Murphy} {et~al.}(2011{\natexlab{b}}){Murphy}, {Condon},
  {Schinnerer}, {Kennicutt}, {Calzetti}, {Armus}, {Helou}, {Turner}, {Aniano},
  {Beir{\~a}o}, {Bolatto}, {Brandl}, {Croxall}, {Dale}, {Donovan Meyer},
  {Draine}, {Engelbracht}, {Hunt}, {Hao}, {Koda}, {Roussel}, {Skibba}, \&
  {Smith}}]{ejm11b}
{Murphy}, E.~J., {Condon}, J.~J., {Schinnerer}, E., {et~al.}
  2011{\natexlab{b}}, \apj, 737, 67

\bibitem[{{Murphy} {et~al.}(2012){Murphy}, {Bremseth}, {Mason}, {Condon},
  {Schinnerer}, {Aniano}, {Armus}, {Helou}, {Turner}, \& {Jarrett}}]{ejm12b}
{Murphy}, E.~J., {Bremseth}, J., {Mason}, B.~S., {et~al.} 2012, \apj, 761, 97

\bibitem[{{Murphy} {et~al.}(1996){Murphy}, {Armus}, {Matthews}, {Soifer},
  {Mazzarella}, {Shupe}, {Strauss}, \& {Neugebauer}}]{twm96}
{Murphy}, Jr., T.~W., {Armus}, L., {Matthews}, K., {et~al.} 1996, \aj, 111,
  1025

\bibitem[{{Niklas} {et~al.}(1997){Niklas}, {Klein}, \& {Wielebinski}}]{nkw97}
{Niklas}, S., {Klein}, U., \& {Wielebinski}, R. 1997, \aap, 322, 19

\bibitem[{{Noeske} {et~al.}(2007){Noeske}, {Weiner}, {Faber}, {Papovich},
  {Koo}, {Somerville}, {Bundy}, {Conselice}, {Newman}, {Schiminovich}, {Le
  Floc'h}, {Coil}, {Rieke}, {Lotz}, {Primack}, {Barmby}, {Cooper}, {Davis},
  {Ellis}, {Fazio}, {Guhathakurta}, {Huang}, {Kassin}, {Martin}, {Phillips},
  {Rich}, {Small}, {Willmer}, \& {Wilson}}]{kgn07}
{Noeske}, K.~G., {Weiner}, B.~J., {Faber}, S.~M., {et~al.} 2007, \apjl, 660,
  L43

\bibitem[{{Onaka} {et~al.}(2007){Onaka}, {Matsuhara}, {Wada}, {Fujishiro},
  {Fujiwara}, {Ishigaki}, {Ishihara}, {Ita}, {Kataza}, {Kim}, {Matsumoto},
  {Murakami}, {Ohyama}, {Oyabu}, {Sakon}, {Tanab{\'e}}, {Takagi}, {Uemizu},
  {Ueno}, {Usui}, {Watarai}, {Cohen}, {Enya}, {Ootsubo}, {Pearson}, {Takeyama},
  {Yamamuro}, \& {Ikeda}}]{to07}
{Onaka}, T., {Matsuhara}, H., {Wada}, T., {et~al.} 2007, \pasj, 59, 401

\bibitem[{{Pannella} {et~al.}(2009){Pannella}, {Carilli}, {Daddi}, {McCracken},
  {Owen}, {Renzini}, {Strazzullo}, {Civano}, {Koekemoer}, {Schinnerer},
  {Scoville}, {Smol{\v c}i{\'c}}, {Taniguchi}, {Aussel}, {Kneib}, {Ilbert},
  {Mellier}, {Salvato}, {Thompson}, \& {Willott}}]{mp09}
{Pannella}, M., {Carilli}, C.~L., {Daddi}, E., {et~al.} 2009, \apjl, 698, L116

\bibitem[{{Sakamoto} {et~al.}(2009){Sakamoto}, {Aalto}, {Wilner}, {Black},
  {Conway}, {Costagliola}, {Peck}, {Spaans}, {Wang}, \& {Wiedner}}]{ks09}
{Sakamoto}, K., {Aalto}, S., {Wilner}, D.~J., {et~al.} 2009, \apjl, 700, L104

\bibitem[{{Salpeter}(1955)}]{salp55}
{Salpeter}, E.~E. 1955, \apj, 121, 161

\bibitem[{{Sanders} {et~al.}(2003){Sanders}, {Mazzarella}, {Kim}, {Surace}, \&
  {Soifer}}]{rbgs03}
{Sanders}, D.~B., {Mazzarella}, J.~M., {Kim}, D.-C., {Surace}, J.~A., \&
  {Soifer}, B.~T. 2003, \aj, 126, 1607

\bibitem[{{Sanders} {et~al.}(1988{\natexlab{a}}){Sanders}, {Soifer}, {Elias},
  {Madore}, {Matthews}, {Neugebauer}, \& {Scoville}}]{dbs88a}
{Sanders}, D.~B., {Soifer}, B.~T., {Elias}, J.~H., {et~al.} 1988{\natexlab{a}},
  \apj, 325, 74

\bibitem[{{Sanders} {et~al.}(1988{\natexlab{b}}){Sanders}, {Soifer}, {Elias},
  {Neugebauer}, \& {Matthews}}]{dbs88b}
{Sanders}, D.~B., {Soifer}, B.~T., {Elias}, J.~H., {Neugebauer}, G., \&
  {Matthews}, K. 1988{\natexlab{b}}, \apjl, 328, L35

\bibitem[{{Schmidt}(1959)}]{ms59}
{Schmidt}, M. 1959, \apj, 129, 243

\bibitem[{{Soifer} {et~al.}(1989){Soifer}, {Boehmer}, {Neugebauer}, \&
  {Sanders}}]{bgs89}
{Soifer}, B.~T., {Boehmer}, L., {Neugebauer}, G., \& {Sanders}, D.~B. 1989,
  \aj, 98, 766

\bibitem[{{Soifer} {et~al.}(2000){Soifer}, {Neugebauer}, {Matthews}, {Egami},
  {Becklin}, {Weinberger}, {Ressler}, {Werner}, {Evans}, {Scoville}, {Surace},
  \& {Condon}}]{bts00}
{Soifer}, B.~T., {Neugebauer}, G., {Matthews}, K., {et~al.} 2000, \aj, 119, 509

\bibitem[{{Stierwalt} {et~al.}(2013{\natexlab{a}}){Stierwalt}, {Armus},
  {Marshall}, {Diaz-Santos}, \& {Charmandaris}}]{ss13b}
{Stierwalt}, S., {Armus}, L., {Marshall}, J., {Diaz-Santos}, T., \&
  {Charmandaris}, V. 2013{\natexlab{a}}, \apj, in preparation

\bibitem[{{Stierwalt} {et~al.}(2013{\natexlab{b}}){Stierwalt}, {Armus},
  {Surace}, {Inami}, {Petric}, {Diaz-Santos}, {Haan}, {Charmandaris}, {Howell},
  {Kim}, {Marshall}, {Mazzarella}, {Spoon}, {Veilleux}, {Evans}, {Sanders},
  {Appleton}, {Bothun}, {Bridge}, {Chan}, {Frayer}, {Iwasawa}, {Kewley},
  {Lord}, {Madore}, {Melbourne}, {Murphy}, {Rich}, {Schulz}, {Sturm}, {U},
  {Vavilkin}, \& {Xu}}]{ss13a}
{Stierwalt}, S., {Armus}, L., {Surace}, J.~A., {et~al.} 2013{\natexlab{b}},
  arXiv:1302.4477

\bibitem[{{U} {et~al.}(2012){U}, {Sanders}, {Mazzarella}, {Evans}, {Howell},
  {Surace}, {Armus}, {Iwasawa}, {Kim}, {Casey}, {Vavilkin}, {Dufault},
  {Larson}, {Barnes}, {Chan}, {Frayer}, {Haan}, {Inami}, {Ishida},
  {Kartaltepe}, {Melbourne}, \& {Petric}}]{vu12}
{U}, V., {Sanders}, D.~B., {Mazzarella}, J.~M., {et~al.} 2012, \apjs, 203, 9

\bibitem[{{Veilleux} {et~al.}(2002){Veilleux}, {Kim}, \& {Sanders}}]{sv02}
{Veilleux}, S., {Kim}, D.-C., \& {Sanders}, D.~B. 2002, \apjs, 143, 315

\bibitem[{{Veilleux} {et~al.}(1995){Veilleux}, {Kim}, {Sanders}, {Mazzarella},
  \& {Soifer}}]{sv95}
{Veilleux}, S., {Kim}, D.-C., {Sanders}, D.~B., {Mazzarella}, J.~M., \&
  {Soifer}, B.~T. 1995, \apjs, 98, 171

\bibitem[{{Veilleux} {et~al.}(1997){Veilleux}, {Sanders}, \& {Kim}}]{sv97}
{Veilleux}, S., {Sanders}, D.~B., \& {Kim}, D.-C. 1997, \apj, 484, 92

\end{thebibliography}

\end{document}